\documentclass[conference]{IEEEtran}
\IEEEoverridecommandlockouts
\usepackage{cite}
\usepackage{amsmath,amssymb,amsfonts}
\usepackage{algorithmic}
\usepackage{graphicx}
\usepackage{float}
\usepackage{subfigure}
\usepackage{textcomp}
\usepackage{xcolor}
\def\BibTeX{{\rm B\kern-.05em{\sc i\kern-.025em b}\kern-.08em
    T\kern-.1667em\lower.7ex\hbox{E}\kern-.125emX}}
\begin{document}

\title{A Real Time Super Resolution Accelerator with Tilted Layer Fusion}

\author{\IEEEauthorblockN{An-Jung Huang$^*$, Kai-Chieh Hsu$^*$ and Tian-Sheuan Chang, \textit{Senior Member, IEEE}}
\IEEEauthorblockA{\textit{Dept. of Electronics Engineering, National Yang Ming Chiao Tung University,} \\
Hsinchu, Taiwan \\
}
\thanks{A.-J. Huang, C.-H. Hsu and T. S. Chang, "A real time super resolution accelerator with tilted layer fusion", ISCAS, 2022 }%
}
\maketitle
\def\thefootnote{*}\footnotetext{These authors contributed equally to this work}\def\thefootnote{\arabic{footnote}}

\begin{abstract}
Deep learning based superresolution achieves high-quality results, but its heavy computational workload, large buffer, and high external memory bandwidth inhibit its usage in mobile devices. To solve the above issues, this paper proposes a real-time hardware accelerator with the tilted layer fusion method that reduces the external DRAM bandwidth by 92\% and just needs 102KB on-chip memory. The design implemented with a 40nm CMOS process achieves 1920x1080@60fps throughput with 544.3K gate count when running at 600MHz; it has higher throughput and lower area cost than previous designs.

\end{abstract}


\section{Introduction}
Deep learning based superresolution (SR) has attracted significant attention recently due to its superior reconstructed image over the traditional methods ~\cite{dong2015image, kim2016accurate,lim2017enhanced}. Many state-of-the-art models~\cite{tong2017image, zhang2018image, xiao2020invertible} have gotten better and better quality, but their large model size and high computational complexity prevent them from real-time hardware implementation. Instead, this paper adopts the hardware-friendly Anchor-based Plain Net (APBN)~\cite{du2021anchor} due to its 8-bit quantized weights, decent PSNR performance (2dB better than the widely used quantized FSRCNN~\cite{dong2016accelerating}), and regular network structure, when compared to other lightweight models~\cite{ahn2018fast, hui2019lightweight, dong2016accelerating}.


Due to the growing interest in applying superresolution in real-time applications, various SR hardware accelerators have been studied. Reference~\cite{kim2018real} adopts the depth-wise convolution for small model size and line-by-line processing for small buffer cost. Reference~\cite{yen2020real} adopts the constant kernel size Winograd convolution for regular hardware design. However, both are designed in a layer-by-layer model execution style which needs to store the layer output to DRAM and load it again for the next layer. It results in high DRAM access amount, especially for high definition image output. Reference~\cite{lee2020srnpu} proposes a selective caching based layer fusion that partitions the input into tiles and uses layer fusion~\cite{alwani2016fused} for processing to avoid intermediate DRAM access. However, it still needs large on-chip memory for tile boundary data in the layer fusion. Block convolution ~\cite{li2021block} ignores these boundary data but will have significant information loss as shown in Fig.~\ref{Information Loss Classical}. Thus, it needs a large tile size and the corresponding large buffer size.



To solve the above problems, we propose an SR accelerator with tilted layer fusion that executes the layer fusion in a tilted addressing between layers. The layer fusion will start its next layer execution as soon as the required input data is ready, which can avoid intermediate data access to the external DRAM due to all these data residing on chip. In practice, the input is usually partitioned into tiles for layer fusion ~\cite{alwani2016fused}, which will need to store the tile boundary data or recompute them to keep the performance intact. The proposed approach will not keep all these boundary data as in ~\cite{alwani2016fused} or discard them as in ~\cite{li2021block} to avoid large buffer size or information loss. Instead, we keep the left and right boundary data but with the tilted layer fusion to reduce the buffer size as well as performance loss, as shown in Fig.~\ref{Information Loss Tilted}. The corresponding hardware uses a simple input broadcasting scheme to reduce control and area overhead. The final results can achieve 1920x1080@60FPS throughput with a much lower area cost compared to previous designs.  



The rest of the paper is organized as follows. Section~\ref{Proposed Tilted Layer Fusion Method} illustrates the proposed tilted layer fusion method. Section~\ref{System Architecture} shows the system architecture of our DLA. The implementation results and comparisons are presented in Section~\ref{Experimental Results}. Finally, this paper is concluded in Section~\ref{Conclusion}.

\section{Proposed Tilted Layer Fusion Method}
\label{Proposed Tilted Layer Fusion Method}

\begin{figure}[t]
\centering
\subfigure[Classical layer fusion~\cite{li2021block}]{
\label{Information Loss Classical}
\includegraphics[width=0.23\textwidth]{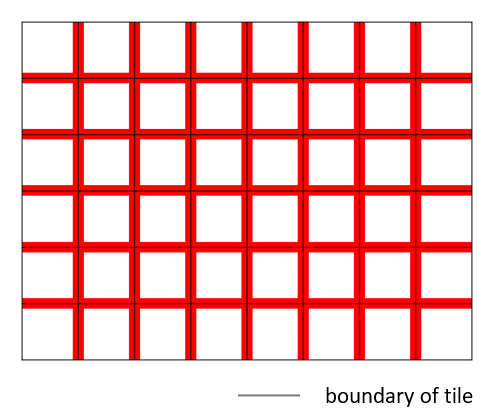}}
\subfigure[Tilted layer fusion]{
\label{Information Loss Tilted}
\includegraphics[width=0.23\textwidth]{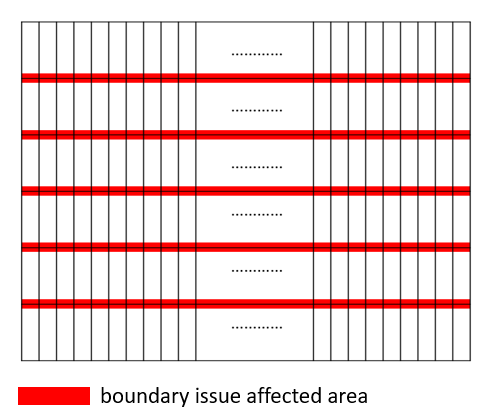}}
\caption{The area affected by recomputation or information loss}
\label{Information Loss}
\end{figure}

\begin{figure}[t]
\centering
\includegraphics[width=0.48\textwidth]{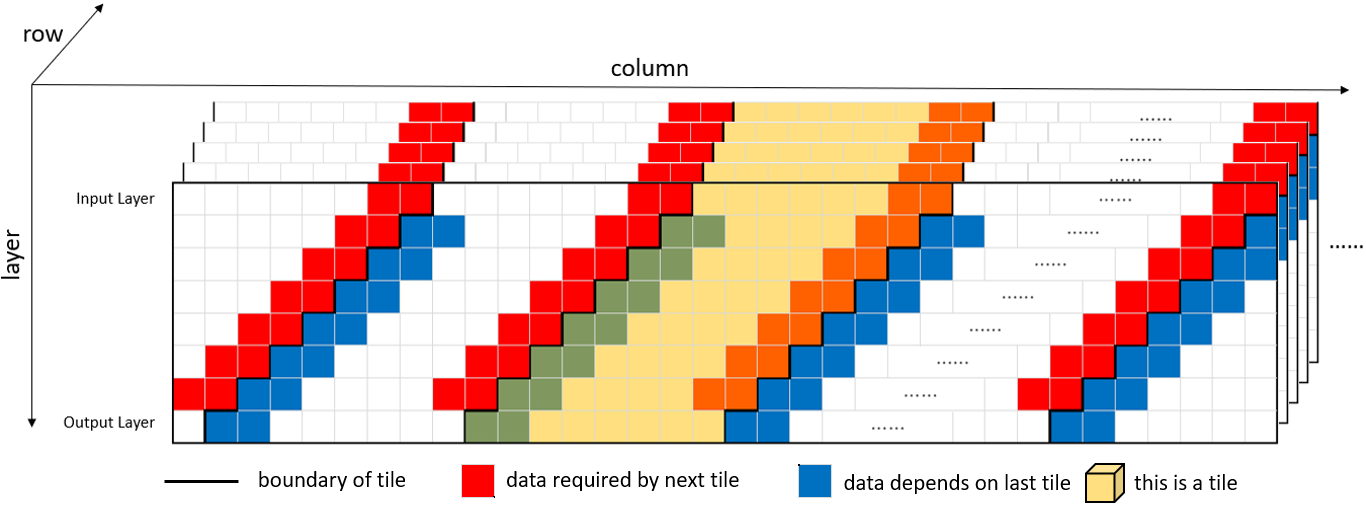}
\caption{Illustration of the tilted layer fusion.}
\label{Tilted Layer Fusion}
\end{figure}

Fig.~\ref{Tilted Layer Fusion} shows a 3D view for all layer inputs by stacking feature maps along the layer axis to demonstrate how the tilted layer fusion works. In which, every block represents a pixel of the feature map in a corresponding layer. Notice that the vertical axis is the layer axis and the horizontal axis is the column axis in a feature map.

As shown in Fig.~\ref{Tilted Layer Fusion}, the feature map input is partitioned into parallelepipedal tiles along the layer instead of rectangular tiles in ~\cite{alwani2016fused} so that the area of a tile on the next layer will be shifted one pixel left. With this, the required data for the next layer at the right boundary (red pixels) will be ready for computation without waiting for other boundary data. 


For the left boundary of the tile (blue pixels), its computation will need data from the left and top neighboring data (red pixels). However, with layer fusion, the left and top neighboring data are just recently finished, which implies we can preserve these data in a limited size buffer until they are not needed anymore.

With the above approach, we can keep the information at the left and right boundaries while only discard information at the top and bottom boundaries, as shown in Fig.~\ref{Information Loss Tilted}. The performance penalty is marginal, less than 0.2dB based on our simulation. Due to this small information loss, the required tile size could be much smaller than previous approaches, and thus it has a smaller buffer size. In this paper, the tile size is selected to be 8x60, and the ignored boundary rows are just 5 rows for the target 640x360 input image as shown in Fig.~\ref{Information Loss Tilted}, which has a negligible impact on the performance penalty.


\section{System Architecture}
\label{System Architecture}

\begin{figure}[t]
\centering
\includegraphics[width=0.48\textwidth]{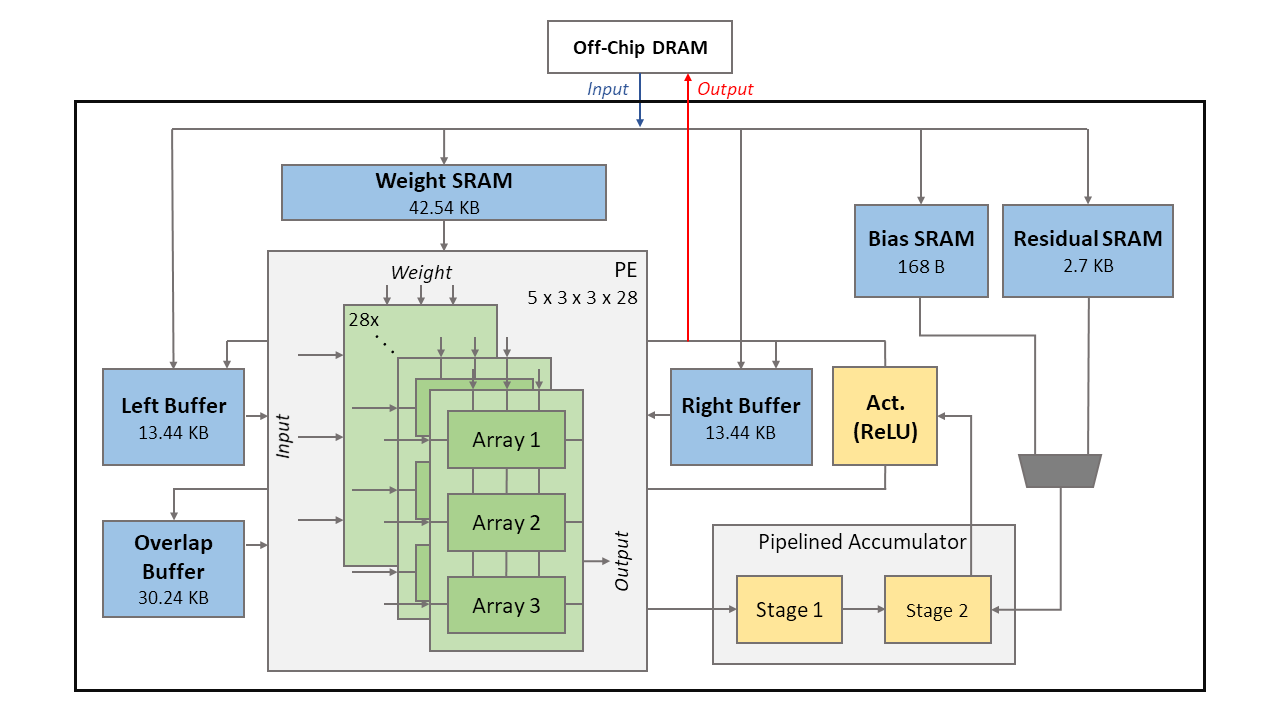}
\caption{System architecture}
\label{System architecture}
\end{figure}

\subsection{Overview}
The adopted model~\cite{du2021anchor} has seven layers. The first six layers are 3x3 convolutions with ReLU activation, and the final layer is a 3x3 convolution followed by a residual-like structure. Based on this model, we propose its hardware architecture as shown in Fig.~\ref{System architecture}. The proposed system architecture consists of 28 PE blocks, 2-stage pipelined accumulator, activation block, ping-pong buffer, overlap buffer, weight SRAM, bias SRAM, and residual SRAM. This design accesses 8-bit input images, weights, and biases from off-chip DRAM and then stores them to the corresponding SRAM buffers.

The whole processing is as follows. First, to process a layer, one of the ping-pong buffers will serve as an input provider to supply input to 28 PE blocks for simultaneously processing 28 channels of input. The output of 28 PE blocks will be sent to the 2-stage pipelined accumulator to complete the convolutions. The results will then go through the activation block for ReLU activation function. Then, the results will be stored in the other ping-pong buffer. Second, to process the next layer, the previous output buffer will serve as input, and the previous input buffer will serve as output. The whole processing is repeated. With this, all intermediate data will be within the chip and save external DRAM access.

\subsection{PE Blocks}

\begin{figure}[t]
\centering
\subfigure[PE array]{
\label{PE array}
\includegraphics[width=0.23\textwidth]{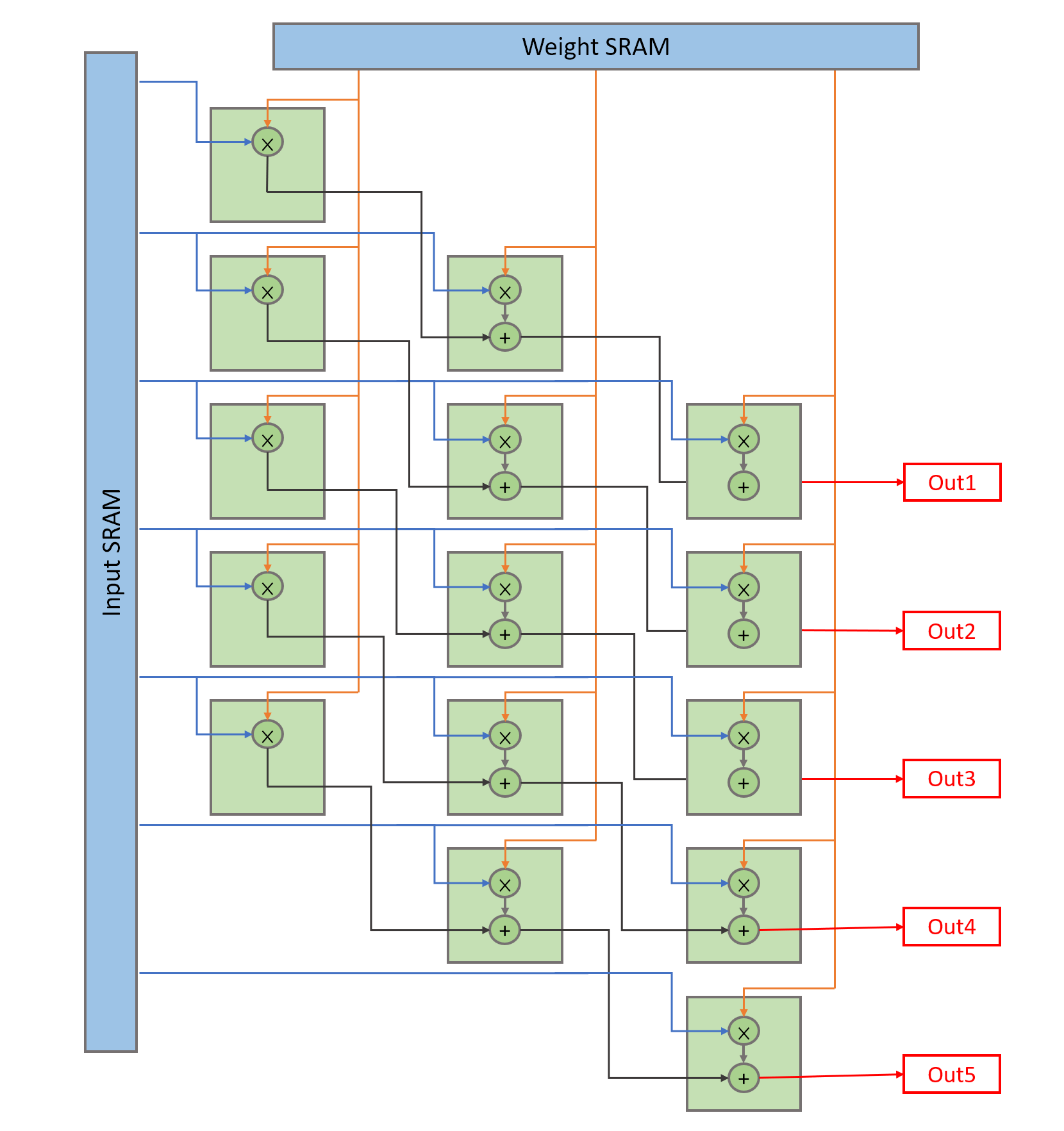}}
\subfigure[Accumulator]{
\label{Accumulator}
\includegraphics[width=0.23\textwidth]{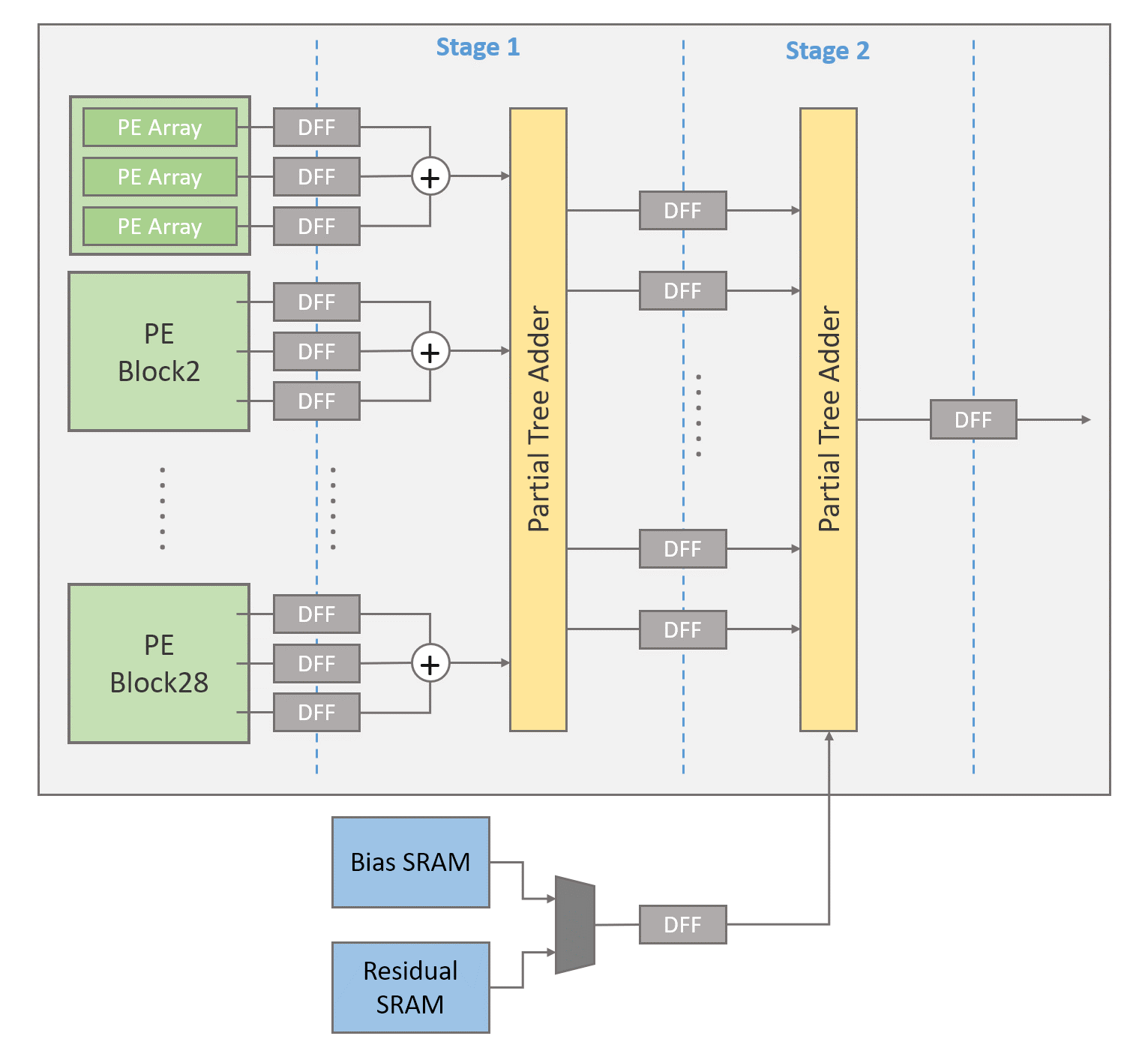}}
\caption{The architecture of PE array and accumulator}
\label{The architecture of PE array and accumulator}
\end{figure}

This design has 28 PE blocks, where each PE block consists of three PE arrays, and a PE array has 5x3 MACs. To ease the arrangement of these MACs, we adopt the input broadcasting as shown in Fig.~\ref{PE array}, where seven input images are broadcasted horizontally, and three weights are broadcasted vertically. These two are multiplied and then added up along the diagonal direction to generate five partial sums for the accumulators.
  
In the adopted model~\cite{du2021anchor}, all the intermediate layers have 28 channels except for the first layer(3 channels) and the final layer(27 channels). By splitting MACs into 28 PE blocks, and letting one PE block process a channel of input at a time, our design can reach an average of $87\%$ hardware utilization with little control overhead. The three PE arrays within one PE block can finish a 3x3 convolution in a cycle, which will be further discussed later.

  

\subsection{Accumulator}

Fig.~\ref{Accumulator} shows the accumulator, which is two-stage pipelined to shorten the critical path. Three partial sums from three PE arrays of the same PE(channel) are added up first, and then the 28 output channels from 28 PE blocks are summed up with a tree adder divided into two partial tree adders. In the second stage of the 2-stage pipelined accumulator, a multiplexer selects whether biases or residuals should be added depending on the current working layer.

\subsection{Data Flow of Convolution}

\begin{figure}[t]
\centering
\includegraphics[width=0.48\textwidth]{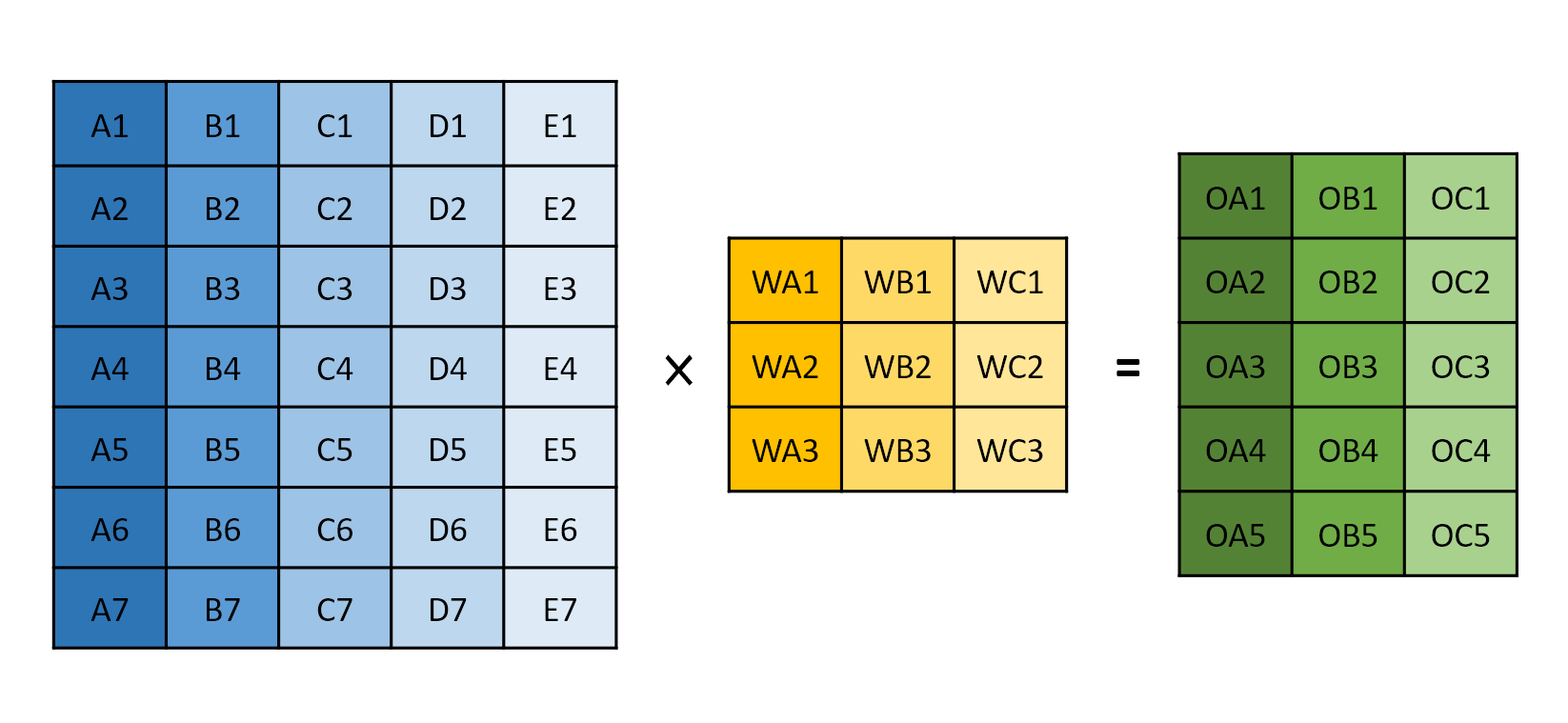}
\caption{A convolution example with 7 × 5 inputs, 3 × 3 weights, and 5 × 3 outputs}
\label{conv example}
\end{figure}

\begin{figure}[t]
\centering
\subfigure[Data flow of a PE block]{
\label{data flow}
\includegraphics[width=0.48\textwidth]{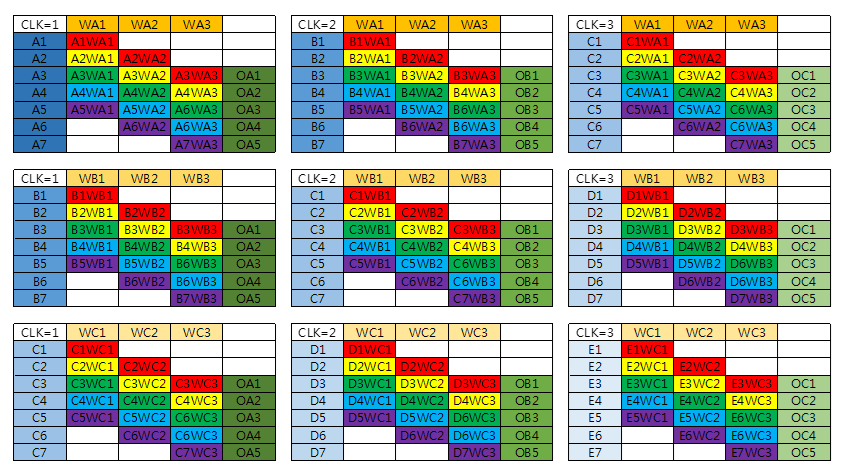}}
\subfigure[Data flow scheduling]{
\label{conv equation}
\includegraphics[width=0.48\textwidth]{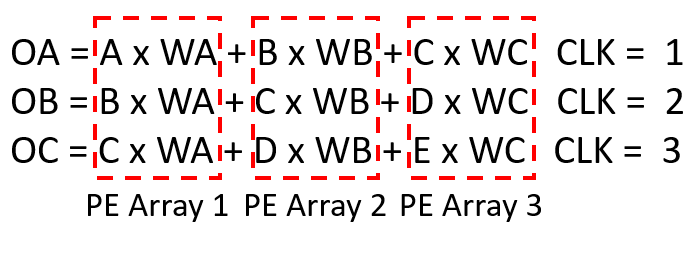}}
\caption{Data flow chart of a PE block processing example in Fig.~\ref{conv example}. The same color products will be added up as output partial sums.}
\label{example data flow}
\end{figure}

Fig.~\ref{data flow} shows the data flow of a PE block process example from Fig.~\ref{conv example}. In our data flow, a column of an input image is broadcasted horizontally while a column of filter weights is broadcasted vertically. With this data flow, the products to be summed up together to complete a convolution are along the diagonal direction. Hence, we use parallelogramical arranged PEs. This design can reduce the control overhead with the expense of allowing two more input images to the PE array.

This vectorwise scheduling is illustrated by the mathematical equations in Fig.~\ref{conv equation}. By sending three consecutive columns of input images and filter weights to three PE arrays, the convolution of one column of output is completed within a cycle. In this case, the example in Fig.~\ref{conv example} only takes three cycles to be completed. To sum up, this data flow achieves high parallelism, simple control, regular structure, and high hardware utilization.

\subsection{I/O Ping-Pong Buffers}

The ping pong buffer stores a tile of data for each layer. When it starts computing a new tile, the system loads the input image data from the external DRAM to the left ping pong buffer. Then, the data are consecutively loaded from the left ping pong buffer to PEs for computation and then stored in the right ping pong buffer after finishing all operations of the current layer. After that, the role of two ping pong buffers is switched to the next layer of computing. This process is repeated layer by layer until finishing the output layer. 

\subsection{Overlap Buffer}
The overlap buffer is a queue-style addressing memory for left and right boundary data. For easy addressing, the current computing layer is regarded as the back layer of the queue, and the last layer is the front layer of the queue. When computing, the data required by the first two columns of the tile are loaded from the front layer, and the results of the last two columns are stored in the back layer. After finishing the computation of a layer, it pops the front layer and the next layer becomes the new back of the queue.

We implement the queue data structure on the overlap buffer by saving the address of the front layer. This addressing helps us calculate the corresponding address of each pixel in both front and back layers in the queue.

\begin{table*}
\caption{Performance Summary and Comparisons with Other Designs}
\begin{tabular}{l|l|l|l|l|l} 
\hline
                          & ~\cite{kim2018real}           & ~\cite{yen2020real}    & ~\cite{chang2018energy}                                                         & SRNPU~\cite{lee2020srnpu}                                                                         & Our Work             \\ 
\hline
SR Method                 & \begin{tabular}[c]{@{}l@{}}DNN\\(1-D CNN)\end{tabular}        & Modified IDN & \begin{tabular}[c]{@{}l@{}}DNN\\(Lightwieght FSRCNN)\end{tabular} & Tile-Based                                                                    & Anchor-Based         \\ 
\hline
Layer Fusion              & None                   & None           & Fused-Layer                                                       & \begin{tabular}[c]{@{}l@{}}Selective Caching Based\\Layer Fusion\end{tabular} & Tilted Layer Fusion  \\ 
\hline
Technology                & \begin{tabular}[c]{@{}l@{}}FPGA\\(Xilinx XCKU040)\end{tabular}& 32 nm        &   \begin{tabular}[c]{@{}l@{}}FPGA\\(Kintex-7410T)\end{tabular}                                              & 65 nm                                                                         & 40 nm                \\ 
\hline
Frequency [MHz]           & 150                  & 200          & 100                                                               & 200                                                                           & 600                  \\ 
\hline
SRAM [KB]                 & 194                  & -           & 945                                                               & 572                                                                           & 102                  \\ 
\hline
Throughput [Mpixels/s]    & 600                  & 124.4        & 520                                                               & 65.9                                                                          & 124.4                \\ 
\hline
Number of Macs            & -                   & 2048         & -                                                                & 1152                                                                          & 1260                 \\ 
\hline
Gate Count                  & -                & 3113.7 k          & -                                                               & -                                                                         & 544.3 k               \\ 
\hline
Normalized Area [$mm^2$] & -                   & -           & -                                                                & 6.06                                                                          & 3.11                 \\ 
\hline
Target Resolution         & 4K UHD (60fps)       & FHD (60 fps) & QHD (120fps)                                                      & FHD (30fps)                                                                   & FHD (60fps)          \\
\hline
\multicolumn{6}{l}{*A 2-input NAND gate is counted as one equivalent gate.}  \\
\multicolumn{6}{l}{*Normalized area is calculated by scaling design to 40nm process.}
\end{tabular}
\label{Comparison}
\end{table*}

\section{Analysis and Experimental Results}
\label{Experimental Results}

\subsection{Buffer Size Analysis}

The buffers in this design serve two purposes. One is for storing the values of feature maps and residuals; the other is for model weights. The following analysis will focus on improving the former since the latter depends on the adopted model. This analysis assumes that the classical layer fusion~\cite{li2021block} uses a 60x60 tile, which is similar to the one used in ~\cite{li2021block} and has the same tile height as our 8x60 tile size. 

In the following, $M_{p}$, $M_{o}$, and $M_{r}$ represent the memory size of ping-pong buffer, overlap buffer, and residual buffer, respectively. $R$ is the number of rows of a tile (length of a tile), $C$ is the number of columns of a tile (width of a tile), $L$ represents the number of layers, and $Ch_{i}$ represents the number of channels of layer $i$; for example, $Ch_{0}$ represents the number of channels of the input layer.

\subsubsection{Ping-Pong Buffers}

As mentioned before, the classical layer fusion method ~\cite{li2021block} cannot have a small tile size due to information loss or recomputation. In this work, the information loss in the horizontal direction has been eliminated by tilted layer fusion, which implies that the tile size restriction on the horizontal direction no longer exists. Therefore, we can reduce the width of the tile as we wish, which helps reduce the buffer size significantly. In the extreme case, the width of the tile can be a single column. In this paper, the width of a tile is selected as 8 columns.

 The size of each ping pong buffer is
\begin{equation}
M_{p}=R\times\:C\:\times\:max\:(Ch_{i})    
\end{equation}

In our case, $C$ is 8 while the classical layer fusion needs to be 60. Our design brings a significant advantage on buffer size cost.

\subsubsection{Overlap Buffer}

In the tilted layer fusion method, the results of one layer have to be reserved for computation of the following layer of the next tile. Therefore, it requires a memory that can contain \emph{number of layer+2} (7+2 for our model)  layers of data. In each layer, 60x2x28 bytes of memory are required to store all data in the last two columns of the tile. The overlap buffer size is
\begin{equation}
M_{o}=L\times\:R\:\times2\:\times\:max\:(Ch_{i})
\end{equation}

Although the proposed method requires additional memory for the overlap buffer, it is still a good trade-off due to memory reduction on ping-pong buffers.

\subsubsection{Residual Buffer}

Because of the tilted layer fusion, the residual buffer has to store $L$ more columns of data. Therefore, the residual buffer size is 
\begin{equation}
M_{r}=Ch_{0}\times\:R\:\times\:(C+L)
\end{equation}

Table~\ref{Size of Buffers} shows the comparison summary of the buffer cost. The proposed approach can save nearly 60\% of the buffer cost. 

\begin{table}
\centering
\caption{Comparison of the buffer size. }
\begin{tabular}{l|l|l} 
\hline
                  & Tilted Layer Fusion & Classical Layer Fusion  \\ 
\hline
Weight Buffer     & 42.54KB             & 42.54KB                 \\ 
\hline
Ping-Pong Buffers & 26.88KB             & 201.6KB                 \\ 
\hline
Overlap Buffer    & 30.24KB             & -                      \\ 
\hline
Residual Buffer   & 2.7KB               & 10.8KB                  \\ 
\hline
Total             & 102.36KB            & 254.94KB                \\
\hline
\end{tabular}
\label{Size of Buffers}
\end{table}

\subsection{Implementation Result and Comparison}

Table~\ref{Comparison} shows the implementation result of the proposed design and the comparison with other designs. This design is synthesized with Synopsys Design Compiler under the TSMC 40nm CMOS process. It achieves 60 fps for x3 scale FHD image generation while running at the 600 MHz clock frequency. The total chip area only occupies 3.11 $mm^{2}$ with 102 KB on-chip SRAM. 

Compared to designs without layer fusion,~\cite{kim2018real} and~\cite{yen2020real}, the required memory bandwidth is significantly reduced. The amount of off-chip DRAM access of this design is reduced from 5.03 GB/sec to 0.41 GB/sec, a total reduction of $92\%$. Thus, even DDR2 DRAM can work well with this design. Compared to designs with layer fusion~\cite{ lee2020srnpu}, this design has a smaller on-chip SRAM due to the adoption of the tilted layer fusion and has a lower area cost.

\section{Conclusion}
\label{Conclusion}
 In this paper, we present an area-efficient hardware accelerator for real-time superresolution applications. It can upscale 640x360 input LR images to 1920x1080 (FHD) HR images at 60 fps. The trade-off between the required memory bandwidth and on-chip memory is resolved by the proposed tilted layer fusion. Compared to previous designs, this design is capable of finishing 60 fps FHD image generation with only 102.36 KB on-chip SRAM while requiring external memory bandwidth is as small as DDR2 transfer rate. The architecture adopts regular vectorwise data flow and diagonalized PE structure to reduce the control overhead. Thus, the presented accelerator can operate at 600 MHz clock frequency and only occupy 3.11 $mm^{2}$ area.

\bibliographystyle{IEEEtran}
\bibliography{bibliography.bib}

\end{document}